\documentstyle[prl,aps,twocolumn]{revtex}
\begin{document}
\draft

\twocolumn[\hsize\textwidth\columnwidth\hsize\csname @twocolumnfalse\endcsname
\title{Dynamical Spontaneous Symmetry Breaking in Quantum Chromodynamics}
\author{Heui-Seol Roh\thanks{e-mail: hroh@nature.skku.ac.kr}}
\address{Department of Physics, Sung Kyun Kwan University, Suwon 440-746, Republic of Korea}
\date{\today}
\maketitle

\begin{abstract}
The longstanding problems of quantum chromodynamics (QCD) as an $SU(3)_C$ gauge
theory, the confinement mechanism and $\Theta$ vacuum, can be resolved by dynamical
spontaneous symmetry breaking (DSSB), which is essential in generating masses of gauge
bosons and hadrons.  The confinement mechanism is the result of massive gluons and the
Yukawa potential provides hadron formation. The evidences for the breaking of discrete
symmetries (C, P, T, CP) during DSSB appear explicitly: baryons and mesons without
their parity partners, the conservation of vector current and the partial conservation
of the axial vector current, the baryon asymmetry $\delta_B \simeq 10^{-10}$, and the
neutron electric dipole moment $\Theta \leq 10^{-9}$.
\end{abstract}

\pacs{PACS numbers:  11.15.Ex, 12.38.-t, 11.15.Tk} ] \narrowtext

Quantum chromodynamics (QCD) \cite{Frit} with quarks and gluons as
fundamental constituents is recognized as the fundamental
dynamical theory for strong interactions. One of the longstanding
problems is however how to manage QCD in the low energy region.
The difficulty in treating QCD at low energy or long range comes
from the fact that the coupling constant becomes so strong that
conventional perturbation theory fails and confinement takes place
in this limit so that the free quark and gluon particles are not
observed. Another longstanding problem of QCD is the $\Theta$
vacuum \cite{Hoof2}, which is a superposition of the various false
vacua, violating CP symmetry.  This paper attempts to
solve the problems nonperturbatively in terms of dynamical
spontaneous symmetry breaking (DSSB) from QCD, to demonstrate that
QND as an $SU(2)_N \times U(1)_Z$ gauge theory for nuclear
interactions originates from QCD as an $SU(3)_C$ gauge theory, and
to propose that QND for nuclear interactions is the analogous
dynamics of the Glashow-Weinberg-Salam (GWS) model as an $SU(2)_L \times U(1)_Y$ gauge
theory \cite{Glas}. The DSSB mechanism is here adopted to strong
interactions characterized by gauge invariance, physical vacuum
problem, and discrete symmetry breaking.
The DSSB mechanism is different from the Higgs mechanism in the GWS model, which has
the problem in generating the fermion mass.
In this scheme, the only free parameter is the strong coupling constant and
several evidences for the violation of discrete symmetries during DSSB are explicitly shown:
baryons and mesons without their parity partners,
the conservation of vector current and the partial conservation of
the axial vector current, the baryon asymmetry $\delta_B \simeq 10^{-10}$ \cite{Stei0},
and the neutron electric dipole moment $\Theta \leq 10^{-9}$ \cite{Alta}.
Furthermore, the mechanism of fermion mass generation and the quantization of intrinsic quantum number
are proposed as consequences of DSSB due to the $\Theta$ vacuum.

The gauge invariant Lagrangian density for QCD with the $\Theta$ vacuum
\cite{Frit,Hoof2,Roh3} is given by
\begin{equation}
\label{qchr}
{\cal L}_{QCD} = - \frac{1}{2} Tr  G_{\mu \nu} G^{\mu \nu}
+ \sum_{i=1}  \bar \psi_i i \gamma^\mu D_\mu \psi_i
+ \Theta \frac{g_s^2}{16 \pi^2} Tr G^{\mu \nu} \tilde G_{\mu \nu},
\end{equation}
where the subscript $i$ stands for the classes of pointlike spinors, $\psi$ for the
spinor, and $D_\mu = \partial_\mu - i g_s A_\mu$ for the covariant derivative with the
strong coupling constant $g_s$. Particles carry the local charges and the gauge
fields are denoted by $A_{\mu} = \sum_{a=0} A^a_{\mu} \lambda^a /2$ with matrices
$\lambda^a$, $a = 0,.., 8$. The field strength tensor is given by $G_{\mu \nu} =
\partial_\mu A_\nu - \partial_\nu A_\mu - i g_s [A_\mu, A_\nu]$
and its dual is $\tilde G_{\mu \nu}$. In the Lagrangian density, the explicit quark
mass term is not contained but the bare $\Theta$ vacuum term is added as a
nonperturbative term to the perturbative term. Since the $G \tilde G$ term is a total
derivative, it does not affect the perturbative aspects of the theory. The $\Theta$
term apparently odd under P, T, and CP operation. An axial current anomaly \cite{Adle}
is taken into account and is linked to the $\Theta$ vacuum in QCD \cite{Hoof2,Roh3}.

The $SU(3)_C$ symmetry for strong force is dynamically spontaneously broken to the
$SU(2)_N \times U(1)_Z$ symmetry and then to the $U(1)_f$ symmetry \cite{Roh3,Roh31}.
The combination of the confinement mechanism and $\Theta$ vacuum explains the DSSB
mechanism in QCD analogous to the Higgs mechanism in electroweak theory
\cite{Glas,Higg}. The Lagrangian density with the $\Theta$ vacuum term possesses all
the known interaction symmetries perturbatively but it does not conserve discrete
symmetries (P, C, T, and CP) . The physical vacuum is not completely symmetric and
DSSB from the normal to physical vacuum takes place. This scheme uses dynamical
symmetric breaking triggering the axial current anomaly \cite{Adle} without
introducing elementary scalar fields. It aims to have DSSB with gauge interactions
alone such as the motivation of technicolor models \cite{Suss}. The concept of DSSB
plays an important role in mass generation in gauge theory which does not have
essentially free parameter.  DSSB consists of two simultaneous mechanisms; the first
mechanism is the explicit symmetry breaking of gauge symmetry, which is represented by
the color factor $c_f$ and the strong coupling constant $g_s$, and the second
mechanism is the spontaneous symmetry breaking of gauge fields, which is represented
by the condensation of color singlet gauge fields.  Gauge fields are generally
decomposed by charge nonsinglet-singlet on the one hand and by even-odd discrete
symmetries on the other hand: they have dual properties in charge and discrete
symmetries.

Four singlet gauge boson interactions in (\ref{qchr}), apart from
nonsinglet gauge bosons, are parameterized by the $SU(3)$ symmetric scalar potential:
\begin{equation}
\label{higs}
V_e (\phi) = V_0 + \mu^2 \phi^2 + \lambda \phi^4
\end{equation}
which is the typical potential with $\mu^2 < 0$ and $\lambda > 0$
for spontaneous symmetry breaking. The first term of the right
hand side corresponds to the vacuum energy density representing
the zero-point energy by even parity singlets. The odd-parity
vacuum field $\phi$ is shifted by an invariant quantity $\langle
\phi \rangle$, which satisfies
\begin{math}
\langle \phi \rangle^2 = \phi_0^2 + \phi_1^2 +  \cdot \cdot \cdot + \phi_{N}^2
\end{math}
with the condensation of odd-parity singlet gauge bosons: $\langle \phi \rangle = (\frac{- \mu^2}{2
\lambda})^{1/2}$. DSSB is relevant for the surface term $\Theta \frac{g_s^2}{16 \pi^2}
Tr G^{\mu \nu} \tilde G_{\mu \nu}$, which explicitly breaks down the
$SU(3)_C$ gauge symmetry for QCD through the condensation of odd-parity singlet gauge
bosons. The $\Theta$ can be assigned by an dynamic parameter by
\begin{math}
\Theta = 10^{-61} \ \rho_G /\rho_m
\end{math}
with the matter energy density $\rho_m$ and the vacuum energy density $\rho_G
= V_e = M_G^4$.

The interaction amplitude in the presence of the gauge boson mass is given by
\begin{equation}
\label{stia}
{\cal M} = - \frac{c_f g_s^2}{4} \frac{1}{k^2 - M_G^2} J^\mu J_{\mu}^{\dagger}
\end{equation}
where the gauge boson mass $M_G$ is inserted in the gauge boson propagator. Parity or
charge conjugation violations due to the condensation of the singlet gauge boson must
be taken into account for current densities $J^\mu = [\bar u \gamma^\mu u]
(c_3^\dagger \lambda^a c_1)$ and $J_{\mu}^{\dagger} = [\bar v \gamma_\mu v]
(c_2^\dagger \lambda_a c_4)$ where $c_i$ with $i=1 \sim 4$ represent local color
charges. The color vector current is conserved (CVC) but the color axial vector
current is partially conserved (PCAC) for strong interactions just as the (V - A)
current is conserved but the (V + A) current is not conserved for weak interactions.
The effective coupling constant at the strong scale is expressed in analogy with the
phenomenological, electroweak coupling constant $G_F =  \frac{\sqrt{2} g_w^2}{8
M_G^2}$ with the weak coupling constant $g_w$:
\begin{equation}
\frac{G_R}{\sqrt{2}} = - \frac{c_f g_s^2}{8 (k^2 - M_G^2)} \simeq \frac{c_f g_s^2}{8 M_G^2}
\end{equation}
where $k$ denotes the four momentum. Similarly and $c_f$ denotes the color factor $c_f$.
The gauge boson mass is generally reduced by the singlet
gauge boson condensation $\langle \phi \rangle$:
\begin{equation}
\label{glma} M_G^2 = M_{H}^2 - c_f g_s^2 \langle \phi \rangle^2 = c_f g_s^2
[A_{0}^2 - \langle \phi \rangle^2]
\end{equation}
where $M_{H} = \sqrt{c_f} g_s A_{0}$ is the gauge boson mass at the grand unification
scale, $A_{0}$ is the singlet gauge boson, and $\langle \phi \rangle$ represents the condensation of
the axial singlet gauge boson. The charge factor $c_f$ used in (\ref{glma}) becomes the
symmetric factor with even parity for singlet gauge boson and is the asymmetric factor
with odd parity for axial singlet gauge boson. The vacuum energy due to the zero-point
energy, represented by the gauge boson mass, is thus reduced by the decrease of the
charge factor and the increase of the axial singlet gauge boson condensation as temperature
decreases.
The essential point is that both the charge coupling constant $c_f \alpha_s$ and the vacuum
expectation value $\langle \phi \rangle$ make the initially massive gauge boson lighter.
The confinement for the charge electric field can be illustrated more rigorously by considering the Yukawa potential \cite{Yuka} due to massive gauge boson.
The Yukawa potential associated with the massive gauge boson is given by
\begin{math}
V (r) = \sqrt{\frac{c_f g_s^2}{4 \pi}} \frac{e^{-M_G (r - l_{QCD})}}{r}
\end{math}
which shows the short range interaction for low energy gauge bosons. Gauge boson
masses are respectively given by $M_{EW} \simeq 10^{2}$ GeV at the weak
scale and $M_{QCD} \simeq 10^{-1}$ GeV at the strong scale. Effective coupling
constants are respectively given by $G_F \simeq 10^{-5} \ \textup{GeV}^{-2}$ at the weak scale
and $G_R \simeq 10^{5} \ \textup{GeV}^{-2}$ at the strong scale.
The effective coupling constant for strong
interactions $G_R$ and Fermi weak coupling constant for weak interactions $G_F$ has
the ratio $G_R/G_F = (\Lambda_{EW} / \Lambda_{QCD})^2 \approx 10^6$.
The gauge boson number density is given by $n_G =
M_G^3$: $n_{EW} \approx 10^{6} \ \textup{GeV}^3 \approx 10^{47} \
\textup{cm}^{-3}$ at the weak scale and $n_{QCD} \approx 10^{-2} \ \textup{GeV}^3
\approx 10^{39} \ \textup{cm}^{-3}$ at the strong scale.

DSSB stages are given by $SU(3)_C \rightarrow SU(2)_N \times U(1)_Z \rightarrow
U(1)_f$ symmetry for strong force. The electric charge quantizations are $\hat Q_e =
\hat I_{3} + \hat Y/2$ for the weak force and $\hat Q_f = \hat C_{3} + \hat Z_c/2$ for
the strong force. The mixing angle for strong interactions $\sin^2 \theta_R$ is the
indication to the DSSB of the $SU(2)_N \times U(1)_Z$ to the $U(1)_f$ gauge symmetry
just as the Weinberg angle for weak interactions $\sin^2 \theta_W$ is the indication
to the DSSB of the $SU(2)_L \times U(1)_Y$ to the $U(1)_f$ gauge symmetry. These
mixing angles relate the strong coupling constant to the coupling for massless gauge
boson dynamics, $\alpha_f = \alpha_n \sin^2 \theta_R = \alpha_n/4$ with the nuclear
coupling constant $\alpha_n$, and the weak coupling constant to the coupling for
photon dynamics, $\alpha_e = \alpha_w \sin^2 \theta_W = \alpha_w/4$ with the weak
coupling constant $\alpha_w$, respectively.  The charge mixing angle is closely
related to massive gauge boson and massless gauge boson generation. In the phase
transition process of the $SU(3)_C \rightarrow SU(2)_N \times U(1)_Z \rightarrow
U(1)_f$ symmetry, there are massive bosons
\begin{math}
A_\mu^{\pm}  =  \frac{1}{\sqrt{2}} ( A_\mu^1 \mp i A_\mu^2)
\end{math}
and
\begin{math}
B^0_\mu  =  \cos \theta_R A_\mu^{3} - \sin \theta_R A_\mu^8 .
\end{math}
The fourth vector orthogonal to $B_\mu$ is identified as massless gauge boson:
\begin{math}
C_\mu = \sin \theta_R A_\mu^{3} + \cos \theta_R A_\mu^8
\end{math}
with the mass $M_C = 0$. Two gauge fields $B_\mu$ and $C_\mu$ are orthogonal
combinations of the gauge fields $A_\mu^{3}$ and $A_\mu^8$ with the mixing angle
$\theta_R$. The generators of this scheme satisfy the relation \cite{Gell}
\begin{equation}
\label{baqu}
\hat Q_f = \hat C_{3} + \hat Z_c/2
\end{equation}
with the longitudinal component of the colorspin operator $\hat C_3$ and the hyper-color charge operator $\hat Z_c$
so that the corresponding current density is presented by
\begin{math}
j_\mu^f = J_\mu^{3} + j_\mu^8/2 .
\end{math}
The interaction in the mixing charge current can provide the relation
\begin{math}
g_f = g_n \sin \theta_R = g_z \cos \theta_R = g_b \cos \theta_R \sin \theta_R
\end{math}
is used.

During phase transition, the discrete symmetries of time reversal (T), parity (P), and
charge conjugation (C) are violated so as to make matter particles massive: since the
product symmetry CPT remains intact, CP symmetry is violated. These violations are
analogous to the nonconservation of the (V + A) current in weak interactions and the
axial vector current in strong interactions. The breaking of discrete symmetries
through the condensation of singlet gauge bosons causes the baryon-antibaryon
asymmetry with $\Theta_{QCD} \simeq 10^{-12}$ at the strong scale. Discrete symmetries
are in general not broken perturbatively but is broken nonperturbatively. The breaking
of discrete symmetries is supported by looking at the observation of pseudoscalar and
vector mesons while their parity partners, scalar and pseudovector mesons, are not
observable at the strong scale; similarly, there is no baryon octet and decuplet
parity pairs. This resolves the $U(1)_A$ problem; the absence of the $U(1)_A$ particle
is due to the nonconservation of the color axial vector current. Singlet gauge bosons
condense in the formation of the hadron and the condensation is relevant for the
partial conservation of axial vector current (PCAC)
\begin{equation}
\partial_\mu J_\mu^5 = \frac{N_f c_f g_s^2}{16 \pi^2} Tr G^{\mu \nu} \tilde G_{\mu \nu}
\end{equation}
with the flavor number $N_f$ and the conservation of vector current (CVC)
$\partial_\mu J_\mu = 0$: this is an example of parity violation. $C$ violation in
baryon as three quark combination is shown in the number difference of the proton and
antiproton as observed in the baryon asymmetry of the present universe; the baryon
asymmetry $\delta_B \approx 10^{-10}$ requires C, T, and CP violations. Based on
observation, there is the tiny CP violation in the nonvanishing electric dipole moment
for the neutron $d_n = 2.7 \sim 5.2 \times 10 ^{- 16} \Theta \ e \ \textup{cm}$
\cite{Alta}, which is reflected by the $\Theta$ vacuum problem $\Theta_{QCD} \leq
10^{-9}$.

Massless gauge bosons (photons) as Nambu-Goldstone (NG) bosons \cite{Namb} are created
during the DSSB. Massless gauge bosons are the quanta of the radiation field that
describes classical light.  Phase transition from $SU(2)_L \times U(1)_Y$ to $U(1)_e$
gauge symmetry and phase transition from $SU(2)_N \times U(1)_Z$ to $U(1)_f$ gauge
symmetry produce massless photons with two transverse polarizations. They are massless
excited modes associated with the generators of the $U(1)_e$ gauge symmetry for weak
force and with the generators of the $U(1)_f$ gauge symmetry for strong force. The
explicit examples of massless gauge bosons for strong force are the intrinsic
vibration modes in nuclear excitation with the typical energies $0.1 \sim 10$ MeV as
noticed by the gamma decays. Massless gauge mode (photon) for the $U(1)_f$ gauge
theory originated from color charges has the coupling constant $\alpha_f = \alpha_s/16
= \alpha_n \sin^2 \theta_R \simeq 1/34$, which is distinct from the coupling constant
$\alpha_e = \alpha_w \sin^2 \theta_W \simeq 1/137$ mediated by the photon for the
$U(1)_e$ gauge theory. However, the massless photon produced by the combination of
color and isospin charges has the coupling constant $\alpha_e \simeq 1/137$ at strong
scale; the conservation of the proton number is the analogy of the conservation of the
electron number. The analogy carriers over to a correspondence between the theory of
electromagnetic radiation in thermal equilibrium and the theory of color radiation in
thermal equilibrium. Each harmonic oscillator of frequency $\omega$ can only have the
energies $(n + 1/2) \omega$, where $n =0,1,2 \cdot \cdot$. This fact leads to the
concept of massless gauge bosons as quanta of the color field whose state is specified
by the number $n$ for each of the oscillators known as massive gauge bosons. Massless
gauge bosons mediate the Coulomb potential ($\sim 1/r$) and the condensation of
singlet gauge bosons makes the confinement potential. The average photon occupation
number in the thermal equilibrium is given by $f_{p} = 1 /(e^{E/T} - 1)$. Massless
photons are quantized by the maximum wavevector mode $N_\gamma \approx 10^{29}$ and
the total photon number $N_{t \gamma} = 4 \pi N_\gamma^3/3 \approx 10^{88}$. The
number density of massless gauge bosons is given by $n_\gamma = 2 \zeta(3) T^3/\pi^3$:
$n_{t \gamma}^{QCD} \approx 10^{36} \ \textup{cm}^{-3}$ at the strong scale.

The fine structure constant $\alpha_s$ for strong interactions is measured
by several experiments \cite{Hinc}:
\begin{math}
\alpha_s (M_Z) \simeq 0.12
\end{math}
at the momentum of the $Z$ boson mass $q = M_Z$
and
\begin{math}
\alpha_s (q) \simeq 0.48
\end{math}
at the momentum of nuclear energy $q \simeq 300$ MeV. Strong coupling constants for
baryons are $\alpha_b = c_f^b \alpha_s = \alpha_s/3$, $\alpha_n = c_f^n \alpha_s =
\alpha_s/4$, $\alpha_z = c_f^z \alpha_s = \alpha_s/12$, and $\alpha_f = c_f^f \alpha_s
= \alpha_s/16$ as symmetric color interactions and $- 2 \alpha_s/3 $, $- \alpha_s/2 $,
$- \alpha_s/6$, and $- \alpha_s/8$ as asymmetric color interactions: $c_f^n = \sin^2
\theta_R$ and $c_f^f = \sin^4 \theta_R$. The color factors introduced are $c^s_f =
(c_f^b, c_f^n, c_f^z, c_f^f) = (1/3, 1/4, 1/12, 1/16)$ for symmetric interactions and
$c^a_f = (-2/3, -1/2, -1/6, -1/8)$ for asymmetric interactions. The symmetric charge
factors reflect intrinsic even parity with repulsive force while the asymmetric charge
factors reflect intrinsic odd parity with attractive force. Asymmetric configuration
for attractive force is confined inside particle while symmetric configuration for
repulsive force is appeared on scattering or decay processes. The color factors
described above are pure color factors due to color charges but the effective color
factors used in nuclear dynamics must be multiplied by the isospin factor $i_f^w =
\sin^2 \theta_W = 1/4$ since the proton and neutron are an isospin doublet as well as
a color doublet. Nucleons as spinors possess up and down colorspins as a doublet just
like up and down strong isospins:
\begin{equation}
{\uparrow \choose \downarrow}_c, \ \uparrow = {1 \choose 0}_c, \ \downarrow = {0
\choose 1}_c .
\end{equation}
This implies that conventional, global $SU(2)$ strong isospin symmetry introduced by Heisenberg \cite{Heis}
is postulated as the combination of local $SU(2)$ colorspin and local $SU(2)$ weak isospin symmetries \cite{Roh31}.
Therefore, the effective color factors are given by
\begin{equation}
c_f^{eff} = i_f^w c_f = i_f^w (c_f^b, c_f^n, c_f^z, c_f^f) = (1/12, 1/16, 1/48, 1/64)
\end{equation}
for symmetric configurations. For example, the electromagnetic color factor for the
$U(1)_f$ gauge theory becomes $\alpha_f^{eff} = \alpha_s/64 \simeq 1/137$ when
$\alpha_s \simeq 0.48$ at the strong scale \cite{Hinc}.

Matter mass generation has features represented by
the $\Theta$ vacuum, dual Meissner effect, and constituent particle mass \cite{Roh3}.
The matter mass is attributed from the dual pairing process due to
dielectric mechanism, which violates the gauge symmetry and discrete symmetries. The
relation between the gauge boson mass and the fermion mass is given by
\begin{equation}
M_G = \sqrt{\pi} m_f c_f \alpha_s \sqrt{N_{sd}}
\end{equation}
where $N_{sd}$ is the difference
number of even-odd parity singlet fermions in intrinsic two-space dimensions. The
above relation stems from the dual pairing mechanism
\begin{math}
M_G = g_{sm}^2 |\psi (0)|^2/ m_f ,
\end{math}
in analogy with electric superconductivity,
where $|\psi (0)|^2 = (m_f c_f \alpha_s)^3$ is the particle probability
density and $g_{sm} = 2 \pi n/\sqrt{c_f} g_s = 2 \pi \sqrt{N_{sd}}/\sqrt{c_f} g_s$ is the
color magnetic coupling constant according to the Dirac quantization condition \cite{Dira}.
The particle number $N_{sd}$ is the difference
number between the number $N_{ss}$ of singlet particles
interacting with charge symmetric configurations and the number
$N_{sc}$ of condensed particles interacting with charge asymmetric
configurations: $N_{sd} = N_{ss} - N_{sc}$.
The fermion mass
formed as the result of confinement mechanism is composed of constituent particles:
\begin{math}
m_f = \sum_i^N m_i
\end{math}
where $m_i$ is the constituent particle mass. In the above, $N$ depends on the
intrinsic quantum number of constituent particles: $N = N_{sd}^{3/2}$. For examples,
$N = 1/B$ with the baryon number $B$ for the constituent quark in the formation of a
baryon and $N = 1/M$ with the meson number $M$ for the constituent quark in the
formation of a meson. A fermion mass term in the Dirac Lagrangian has the form $m_f
\bar \psi \psi = m_f (\bar \psi_A \psi_V + \bar \psi_V \psi_A)$ where the mass term is
equivalent to a helicity flip. Vector fermions are put into $SU(2)$ doublets and
axial-vector ones into $SU(2)$ singlets.  In the dual pairing mechanism, discrete
symmetries P, C, T, and CP are dynamically broken due to massive gauge bosons
\cite{Roh3}. Electric monopole, magnetic dipole, and electric quadrupole remain in the
matter space but magnetic monopole, electric dipole, and magnetic quadrupole condense
in the vacuum space as the consequence of P violation. Antibaryon particles condense
in the vacuum space while baryon particles remain in the matter space as the result of
C and CP violation at the strong scale: the baryon asymmetry $\delta_B \simeq
10^{-10}$ \cite{Stei0}. The electric dipole moment of the neutron and no parity
partners in hadron spectra are the typical examples for P, T, and CP violation at the
strong scale. The fine or hyperfine structures of a fermion mass include spin-spin,
isospin-isospin, and colorspin-colorspin interactions due to intrinsic angular momenta
of $SU(2)$ gauge theories:
\begin{equation}
\triangle E \propto \alpha_m \frac{\vec s_i \cdot \vec s_j}{m_i m_j} |\psi (0)|^2
+ \alpha_i \frac{\vec i_i \cdot \vec i_j}{m_i m_j} |\psi (0)|^2
+ \alpha_s \frac{\vec \zeta_i \cdot \vec \zeta_j}{m_i m_j} |\psi (0)|^2
\end{equation}
where $\alpha_m$, $\alpha_i$, and $\alpha_s$ are respectively coupling constants. The
approach can be applied to investigate the masses of hadrons, which justify the
constituent quark model as an effective model of QCD at low energies from this
viewpoint. If interactions due to colorspin and isospin contributions are absorbed to
the constituent quark mass, the meson mass of the conventional constituent quark model
is obtained:
\begin{equation}
m_m = m_1 + m_2 + A \frac{\vec \sigma_1 \cdot \vec \sigma_2}{m_1 m_2}
\end{equation}
where $\vec \sigma_1 \cdot \vec \sigma_2 = 4 \vec s_1 \cdot \vec s_2 =1$ for vector mesons and
$\vec \sigma_1 \cdot \vec \sigma_2 = -3$ for pseudoscalar mesons are given
and $A = \frac{8 \pi g_s^2 |\psi (0)|^2}{9}$.
By the same token, the baryon mass of the conventional constituent quark model is obtained:
\begin{equation}
m_b = m_1 + m_2 + m_3 + A' \sum_{i>j} \frac{\vec \sigma_i \cdot \vec \sigma_j}{m_i m_j}
\end{equation}
where $A' = \frac{4 \pi g_s^2 |\psi (0)|^2}{9}$.
Since $\sum \sigma_i \cdot \sigma_j = 4 s_i \cdot s_j = 2 [s(s+1) - 3s(s+1)]$ with the total spin
$\vec S = \vec s_1 + \vec s_2 + \vec s_3$,
$\sum \vec \sigma_i \cdot \vec \sigma_j = 3$ for decuplet baryons and
$\sum \vec \sigma_i \cdot \vec \sigma_j = -3$ for octet baryons are given.
The constituent quark model illustrates reasonable agreement in hadron spectra within a few percent deviation.

This study proposes that the longstanding problems of quantum chromodynamics (QCD) as
an $SU(3)_C$ gauge theory, the confinement mechanism and $\Theta$ vacuum, can be
resolved by dynamical spontaneous symmetry breaking (DSSB) through the condensation of
singlet gluons and quantum nucleardynamics (QND) as an $SU(2)_N \times U(1)_Z$ gauge
theory is produced. DSSB in QCD is essential in generating masses of gauge bosons and
hadrons. During DSSB, common features in gauge theories appear: massive gauge boson,
massless gauge boson (photon) as the NG boson, nonperturbative spontaneous breaking of
discrete symmetries, charge mixing angle, and coupling constant hierarchy. $G_R$ is
realized as the effective coupling constant for a massive gluon, $G_R /\sqrt{2} = c_f
g_s^2/8 M_G^2 \approx 10 \ \textup{GeV}^{-2}$ with the gauge boson mass $M_G = M_{QCD}
\approx 10^{-1}$ GeV, the strong coupling constant $g_s$, and the color factor $c_f$.
The condensation of the singlet gauge field $\langle \phi \rangle$ triggers the
current anomaly and subtracts the gauge boson mass, $M_G^2 = M_{H}^2 - c_f g_s^2
\langle \phi \rangle^2 = c_f g_s^2 [A_{0}^2 - \langle \phi \rangle^2]$, as the vacuum
energy. Spontaneous symmetry breaking in gauge theories is in this work extended to
dynamical spontaneous symmetry breaking mechanism, which has only one free coupling
constant, to generate gauge boson masses and fermion masses as well as to explain
nonperturbative discrete symmetry breaking in strong interactions.

\end{document}